# Anomalous magnetotransport in wide quantum wells

J. Oswald, G.Span, A. Homer, G. Heigl, P. Ganitzer

Institute of Physics, University of Leoben, Franz Josef Str. 18, A-8700 Leoben, Austria

D.K. Maude, J.C. Portal

High Magnetic Field Laboratory, CNRS, 25 Avenue des Martyrs, BP 166 Grenoble, France

*We present magneto transport experiments of quasi 3D PbTe wide quantum wells. A plateau-like structure in the Hall resistance is observed, which corresponds to the Shubnikov de Haas oscillations in the same manner as known from the quantum Hall effect. The onsets of plateaux in $R_{xy}$ do not correspond to 2D filling factors but coincide with the occupation of 3D (bulk-) Landau levels. At the same time a non-local signal is observed which corresponds to the structure in $R_{xx}$ and $R_{xy}$. and fulfils exactly the Onsager-Casimir relation ($R_{ij,kl}(B) = R_{kl,ij}(-B)$). We explain the behaviour in terms of edge channel transport which is controlled by a permanent backscattering across a system of "percolative EC - loops" in the bulk region. Long range potential fluctuations with an amplitude of the order of the subband splitting are explained to play an essential role in this electron system.*

Keywords: semiconductors, quantum wells, electronic transport, quantum Hall effect

**1. Introduction:**

Recently observed large amplitude fluctuations in local and non-local magneto transport experiments [1,2] have lead to the idea that a parallel contribution of a bulk-like electron system and an edge channel (EC) system can cause these experimental observations [3,4]. The main idea is that in PbTe there are two different electron systems with significantly different subband splitting. In a suitable wide quantum well we can get an intermediate regime where the electron system with the large effective mass can lead already to a dissipative bulk-like conduction while the electron system with the small confining mass can lead to nearly dissipation less EC-conduction. On this background the samples have been described by an equivalent circuit which represents the bulk-like system by an Ohmic resistor network and the EC-system by ideal dissipation-less EC-conduction. The electron channel width of the PbTe WQW's is of the order of several 100 nm. This leads to a typical subband splitting of the order of 1..2 meV for the small mass and about a factor of 10 less for the large effective mass. Native potential fluctuations, which result from the statistical distribution of the donors, are also of the order of 1...3 meV. Due to the high dielectric constant of PbTe ($\varepsilon \approx 1000$), these potential fluctuations are expected to be smooth long-range fluctuations with a lateral size of the order of 100nm and more. Consequently they are about an order of magnitude larger than the cyclotron radius at magnetic fields around 10 Tesla. Therefore the broadening of the LL's due to potential fluctuations can also be understood to be just a lateral variation of the LL energy. From this point of view it is likely to expect a network of well developed "percolative" EC-loops in the interior of the sample which follow the equipotential lines of the potential fluctuations[5]. Although separated from each other, they may provide some probability for transport via resonant tunnelling. Depending on the density of these loops, a tunnelling process across the loops is a likely mechanism for back scattering of edge electrons. The observed conductance fluctuations have been attributed to this back scattering process, which becomes most dominant in narrow contact leads [3,4].

In this paper we present new experimental results, which don't focus on the fluctuations but focus on the structure of the magneto transport data it self. We demonstrate that in WQW's with a high bulk doping level of the electron channel a structure appears in $R_{xy}$, which has strong similarities with onsets of plateaux of the quantum Hall effect. Indeed, the onsets of the plateau like features in $R_{xy}$ correspond to positions where $R_{xx}$ produces minima with respect to a linear slope of the background.

## 2. Experiments:

The experiments have been performed in a super conducting magnet in fields up to B=17 T with a dilution refrigerator at temperatures down to T = 40mK. Selective contacts to the embedded electron channel are made by evaporating Indium and subsequent diffusion [6]. A standard lock-in technique was used (f = 13.7 Hz) and the constant ac-current was chosen to be ≤100 nA. The samples have been measured in local and non-local contact configuration. Systematic experiments with different sample geometry and doping levels of the electron channel have been performed. The most interesting data have been obtained at a high bulk-doping level of the electron channel.

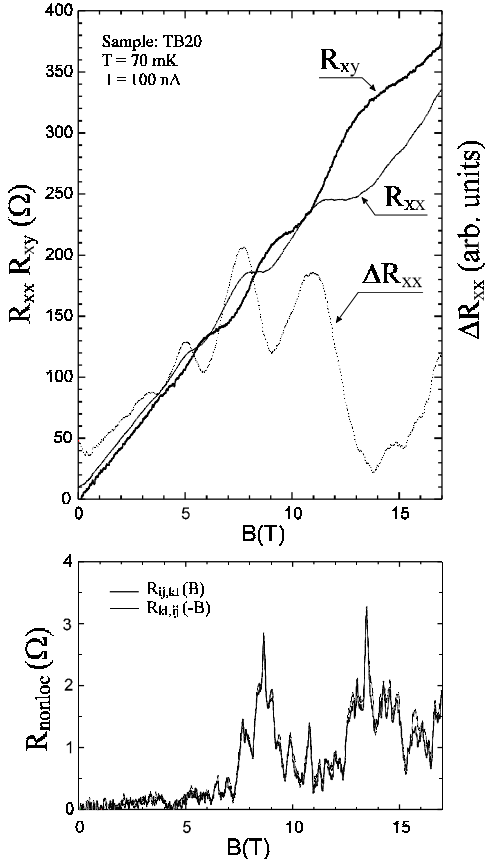

Figure 1: *Magneto transport data of sample TB20. The upper diagram shows $R_{xx}$ and $R_{xy}$ data and the lower diagram shows the corresponding non-local data.*

Figure 1 shows $R_{xx}$, $R_{xy}$ and non-local Data. In addition, the $R_{xx}$ data are shown after subtracting the linear back ground (labelled $\Delta R_{xx}$). Bulk-Shubnikov de Haas (SdH) oscillations can be observed up to B=15 T. The resulting bulk density in the electron channel is $6 \times 10^{17} cm^{-3}$, which results in an effective electron channel width of about 500nm.

## 3. Discussion:

The basis for the discussion are the following experimental facts:
(i) There are onsets of plateaux in $R_{xy}$ which correspond to minima in $\Delta R_{xx}$.
(ii) The maxima in $\Delta R_{xx}$ appear in the regime of maximal slope in $R_{xy}$.
(iii) The "plateaux" correspond to the bulk-SdH oscillations and therefore cannot be attributed to 2D-fillingfactors.
(iv) the non-local signal including the superimposed fluctuation like structure fulfil exactly the Onsager-Casimir relation.

Already in [3,4] the existence of potential fluctuations has been explained to play a key-role in the WPQW's. Therefore we consider the possible influence of long range potential fluctuations in a WQW first.

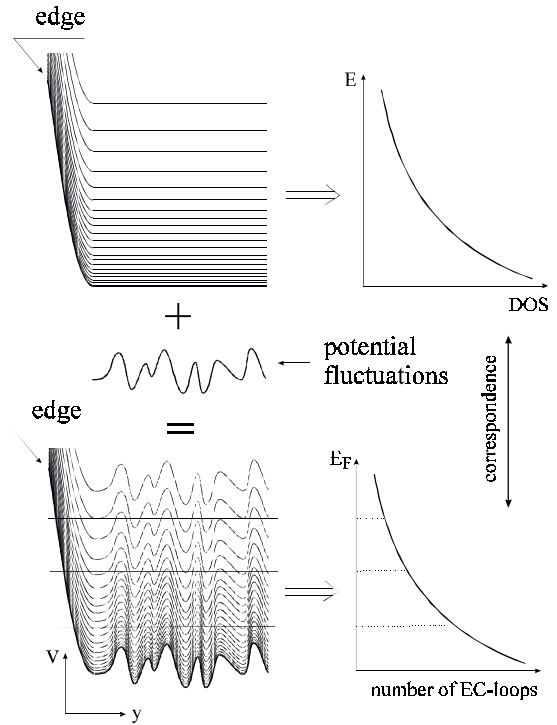

*Figure 2: Schematic representation of the influence of potential fluctuations.*

The upper left part of *Figure 2* shows the Landau subband levels without potential fluctuations. Since in a WQW the subband splitting is much smaller than the LL-splitting, the level sequence is determined by the subband spectrum. The energy spectrum therefore is a single Landau level which is split by the spectrum of a wide square potential well. The overall DOS on the upper right part of *Figure 2* results from the contributions of the individual (broadened) Landau peaks associated with the energy levels [1]. The lower right part *Figure 2* shows the effect of potential fluctuations. Although the level spectrum results from the subbands, the levels are physically Landau levels and are therefore shifted in parallel with the local potential of the long range potential fluctuations. If we consider the Fermilevel at different energy positions, one can see that the number of intersections with subband LL's depends on the Fermienergy. These intersections result in percolative EC-loops in the bulk region. Therefore the number of loops and/or the loop density will qualitatively depend on the Fermienergy like indicated in the lower right part of *Figure 2*. It is striking, that one gets qualitatively a correspondence between the 3D-DOS in the upper right part and the number of EC-loops in the bulk region in the lower right part of *Figure 2*. The density of the created EC-loops will be a maximum if the Fermi level is at low energies near the subband ground state. Since the subband LL's are sitting on top of a bulk LL, the subband ground state corresponds to a bulk LL. Consequently we get a maximal loop density, if the Fermilevel crosses a bulk LL. In contrast, the loop density is minimal if the Fermilevel is high above the bulk LL. From this point of view the 3D-density of states maps directly into the density of created magnetic bound states in the bulk region. The coupling between opposite edges across these loops will depend on the density of these loops. A large loop density will lead to a more effective coupling than a low loop density. A small subband splitting leads to a high density of magnetic bound states and thus a high dissipation due to EC-back scattering occurs in this system. This argument can be applied in principle also to the dissipative bulk-like system. Since in general dissipation must show up as a finite Ohmic longitudinal resistance ($R_{xx}$), it is clear that the Ohmic part of the total behaviour will be dominated by the most dissipative system, which will be that one with the smallest subband-splitting. In this way we arrive at the already mentioned equivalent circuit. So far it is not important, if the dissipative bulk-like system is a system with true bulk conduction or if it is an EC-system with significantly stronger back scattering than in the other system.

In a sweeping magnetic field the Fermi level moves relative to the LL's. As one can see from *Figure 2*, the loop density is maximal if the Fermilevel is near the subband ground state which is at the same time the energy of a bulk LL. Therefore one can expect maximal dissipation and hence a maximum of $R_{xx}$, if the Fermilevel intersects with a bulk LL. In this context we would like to point out a major difference to standard 2D systems: In 2D systems, the disorder potential changes with the position of the Fermi energy and it is maximal if the Fermi level is in between two LL's. This happens because the screening of potential fluctuations depends on the number of mobile carriers. Consequently the potential fluctuations are strongest if the Fermi level is between two LL's where there are only localized states. This should lead to an opposite effect as explained above. However, in the PbTe WQW's we have two electron systems and the screening is permanently maintained by the 3D-like electrons which cannot build localised states. Consequently we can expect that the trend, which we obtain from considering the subband splitting at different positions of the Fermi energy, is not cancelled by a Fermi level dependent screening of the potential fluctuations.

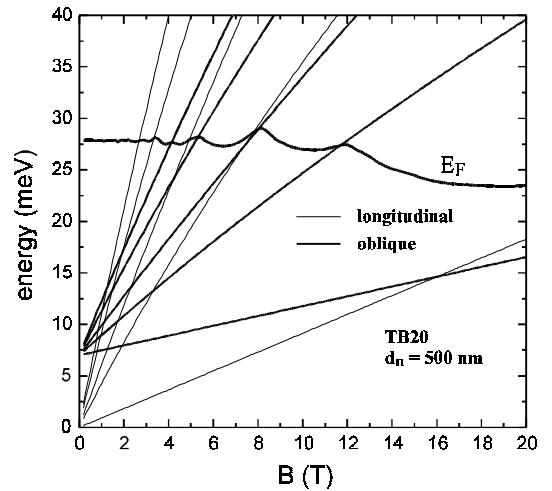

*Figure 3 Landauniveaus and Fermienergy vs. magnetic field for parameters according to sample TB20*

Figure *3* shows the calculation of the LL's and the Fermienergy as a function of magnetic field for the parameters of the investigated sample. There are 3 dominating intersections of the Fermienergy with bulk LL's, which correspond very well to the experimental data of Figure 1. It is easily seen, that at magnetic fields, where $E_F$ approaches a bulk LL, there is a maximum in $\Delta R_{xx}$ and a maximal slope in $R_{xy}$. If $E_F$ drops below a bulk LL, the DOS and thus the loop density can be expected to be minimal. Indeed, in this regime a minimum in $\Delta R_{xx}$ and an onset of a plateau in $R_{xy}$ appears. If we look at the non-local signal, we realise, that it is at the maximum if $\Delta R_{xx}$ is at the minimum. But also this is easily explained, because non-local transport will be most efficient, if EC-back scattering is minimal. The minima in $R_{xx}$ indicate that back-scattering is minimal and thus the non-local transport can be maximal, exactly consistent with the experimental observation.

The discussion of $R_{xy}$ is a little more complicated because also the magnetic field dependent carrier transfer between both systems has to be considered. In general the total filling factor of the complete electron system decreases smoothly, proportional to the reciprocal magnetic field. If there would be just a single electron system and if there are no localized states like in our case, $R_{xy}$ would raise up exactly in a straight line. Since we have two electron systems, the oscillating density of states at the Fermilevel in both systems leads to a back and forth carrier re-distribution in order to keep the total filling factor monotone. If the Fermi level is moving in the low DOS regime of the 2D-like system, the Fermi level pinning is dominated by the background DOS of the 3D-like electron system. This leads basically to a carrier re-distribution to the 2D-like system and hence the further decrease of the fillingfactor of the 2D-like electron system is slowed down some what. Since in this regime there is minimal dissipation in the 2D-like system, the Hall voltage is dominated by this system and therefore also the further increase of the Hall voltage slows down some what and thus produces an onset of a plateau-like feature.

The QHE-like behaviour of our system is easily understood, if one compares our system with a standard QH system. We find a remarkable correspondence between our quasi 3D system and a standard 2D-QH-system:
(i) In standard QH-systems EC-back scattering occurs if a 2D-LL crosses $E_F$ while in the WPQW EC-back scattering is maximal if $E_F$ crosses a bulk-LL.
(ii) The energy range between bulk-LL's in the WPQW corresponds to the energy range between 2D- LL's in the QH-system. While in this regime EC-backscattering is completely suppressed in the standard QH-system, it is just significantly reduced in the WPQW-system.
(iii) If the Fermilevel in 2D systems moves in a so called DOS-gap, the Fermilevel is pinned by localized states, which allows a continuous change of the total fillingfactor. In our quasi 3D system the role of the localized states is taken over by the bulk-like system, which, however, does not create localised states but realises a less conductive system with much dissipation. Therefore the zeros in $R_{xx}$ are missing and a linear slope is superimposed to $R_{xx}$ instead. Such a linear increase of the magneto resistance with magnetic field has been also the subject of investigations in other systems [7]. An explanation by Büttiker [8] is based on EC-conduction in the presence of lateral inhomogeneities in a quasi-3D electron channel. This is in principle consistent with our explanation where the inhomogeneities are represented by the long range potential fluctuations. A further consideration of the linear magneto resistance in PbTe WPQW's is the subject for publication else where.

Finally the non-local signal clearly satisfies the Onsager-Casimir relation for the complete pattern. This is another indication that the general non-local behaviour and the superimposed fluctuations have the same physical origin.

Based on these arguments one can conclude, that the physical mechanisms which lead to the well known QHE in 2D systems, dominate also the magneto transport properties of high mobility WPQW- systems. Consequently the experiments shown in Figure 1 can be seen as the first observation of a 3D-QHE in WPQW's.

**4. Summary :**

We have presented magnetotransport experiments which show onsets of plateaux in $R_{xy}$ which behave similar like known from the QHE of 2D systems. In contrast, the position of the plateaux is correlated with the occupation of 3D Landau levels and does not correspond to the 2D filling factor. We have been able to explain this behaviour in terms of edge channel transport, which is subject of a permanent back scattering. The back scattering mechanism is explained to be controlled by "percolative edge channels" in the bulk region, which result from smooth, long range potential fluctuations with an amplitude of the order of the subband splitting. On this background we conclude that the transport may already be strongly influenced by "quantum percolation".

Financial support by Fonds zur Förderung der wissenschaftlichen Forschung (FWF) Austria, Project: P10510 NAW.